\documentclass[aip,preprint]{revtex4-1}
\usepackage{graphicx}
\usepackage{dcolumn}
\usepackage{bm}
\usepackage{color}
\usepackage{tabularx}
\usepackage{array}
\usepackage{amsmath}
\usepackage{pbox}
\usepackage{stmaryrd}  

\begin{document}

\title{\Large Probing magnon-magnon coupling \textcolor{black}{in exchange coupled Y$_3$Fe$_5$O$_{12}$/Permalloy bilayers} with magneto-optical effects}


\author{Yuzan Xiong}
\affiliation{Department of Physics, Oakland University, Rochester, MI 48309, USA}
\affiliation{Department of Electronic and Computer Engineering, Oakland University, Rochester, MI 48309, USA}

\author{Yi Li}
\email{yili@anl.gov}
\affiliation{Materials Science Division, Argonne National Laboratory, Argonne, IL 60439, USA}

\author{Mouhamad Hammami}
\affiliation{Department of Physics, Oakland University, Rochester, MI 48309, USA}

\author{Rao Bidthanapally}
\affiliation{Department of Physics, Oakland University, Rochester, MI 48309, USA}

\author{Joseph Sklenar}
\affiliation{Department of Physics and Astronomy, Wayne State University, Detroit, MI 48201, USA}

\author{Xufeng Zhang}
\affiliation{Center for Nanoscale Materials, Argonne National Laboratory, Argonne, IL 60439, USA}

\author{Hongwei Qu}
\affiliation{Department of Electronic and Computer Engineering, Oakland University, Rochester, MI 48309, USA}

\author{Gopalan Srinivasan}
\affiliation{Department of Physics, Oakland University, Rochester, MI 48309, USA}

\author{John Pearson}
\affiliation{Materials Science Division, Argonne National Laboratory, Argonne, IL 60439, USA}

\author{Axel Hoffmann}
\affiliation{Department of Materials Science and Engineering, University of Illinois at Urbana-Champaign, Urbana, IL 61801, USA}
\affiliation{Materials Science Division, Argonne National Laboratory, Argonne, IL 60439, USA}

\author{Valentine Novosad}
\affiliation{Materials Science Division, Argonne National Laboratory, Argonne, IL 60439, USA}

\author{Wei Zhang}
\email{weizhang@oakland.edu}
\affiliation{Department of Physics, Oakland University, Rochester, MI 48309, USA}
\affiliation{Materials Science Division, Argonne National Laboratory, Argonne, IL 60439, USA}


\date{\today}

\begin{abstract}

We demonstrate the magnetically-induced transparency (MIT) effect in Y$_3$Fe$_5$O$_{12}$(YIG)/Permalloy(Py) coupled bilayers. The measurement is achieved via \textcolor{black}{a heterodyne} detection of the coupled  magnetization dynamics using a single wavelength that probes the magneto-optical Kerr and Faraday effects of Py and YIG, respectively. Clear features of the MIT effect are evident from the deeply modulated ferromagnetic resonance of Py due to the perpendicular-standing-spin-wave of YIG. We develop a phenomenological model that nicely reproduces the experimental results including the induced amplitude and phase evolution caused by the magnon-magnon coupling. Our work offers a new route towards studying phase-resolved spin dynamics and hybrid magnonic systems. 

\end{abstract}

\maketitle

\section{Introduction}

Hybrid magnonic systems are becoming rising contenders for coherent information processing \cite{nakamura_apex2019,hu_ssp2018,bhoi_ssp2019}, owing to their capability of connecting distinct physical platforms in quantum systems as well as the rich emerging physics for new functionalities \cite{nakamura_science2015,yili_prl2019,luqiao_prl2019,bai_prl2015,huebl_prl2013,xufeng_prl2014,xufeng_ncomm2015,bauer_prb2015,harder_prl2018,bhoi_prb2019,thompson_prl2017,you_ncomm2017,tabuchi_prl2014,kakkawa_prl2016,xzhang_sciadv2016,bai_prl2017,andrich_npjq2017,haigh_prb2019}. Magnons have been demonstrated to efficiently couple to cavity quantum electrodynamics systems including superconducting resonators and qubits \cite{huebl_prl2013,nakamura_science2015,haigh_prb2019,yili_prl2019,luqiao_prl2019}; magnonic systems are therefore well-positioned for the next advances in quantum information. In addition, recent studies also revealed the potential of magnonic systems for microwave-optical transduction \cite{osaka_prl2016,xzhang_prl2016,hisatomi_prb2016,haigh_prl2016,rasing_rmp2010,jieli_prl2018,dany_science2020}, which are promising for combining quantum information, sensing, and transduction.  

To fully leverage the hybrid coupling phenomena with magnons, strong and tunable couplings between two magnonic systems have attracted considerable interests recently \cite{weiler_prl2018,yu_prl2018,qin_srep2018,yili_magnon2019}. They can be considered as hosting hybrid magnonic modes in a ``magnonic cavity" as opposed to microwave photonic cavity in cavity-magnon polaritons (CMPs) \cite{nakamura_apex2019,hu_ssp2018,bhoi_ssp2019}, which allows excitations of forbidden modes and high group velocity of spin waves owing to the state-of-the-art magnon bandgap engineering capabilities \cite{yu_prl2018,kruglyak_jphysd2010}. The detuning of the two magnonic systems can be easily engineered by the thickness of the thin films, which set the wavenumbers and the corresponding exchange field. Furthermore, in such strongly coupled magnetic heterostructures, both magneto-optical Kerr and Faraday effects can be utilized for light modulation, in terms of light reflection by metals and/or transmission in
insulators, respectively. In this architecture, the freedom of lateral dimensions is maintained for device fabrication and large-scale, on-chip integration.

To date, both magnon-photon and magnon-magnon couplings are predominantly investigated by the cavity ferromagnetic resonance (FMR) spectroscopy, i.e. microwave transmission and/or reflection measurements, typically involving a vector-network analyzer (VNA) or a microwave diode \cite{nakamura_science2015,yili_prl2019,luqiao_prl2019,bai_prl2015,huebl_prl2013,xufeng_prl2014,xufeng_ncomm2015,bauer_prb2015,weiler_prl2018,yu_prl2018,qin_srep2018,yili_magnon2019,an_prb2020}. Strong magnon-magnon couplings have been observed in yttrium iron garnet (Y$_3$Fe$_5$O$_{12}$, YIG) coupled with ferromagnetic (FM) metals, where exchange spin waves were excited by a combined action of exchange, dampinglike, and/or fieldlike torques that are localized at the interfaces \cite{weiler_prl2018,yu_prl2018,qin_srep2018,yili_magnon2019}.

\begin{figure*}[htb]
 \centering
 \includegraphics[width=6.5 in]{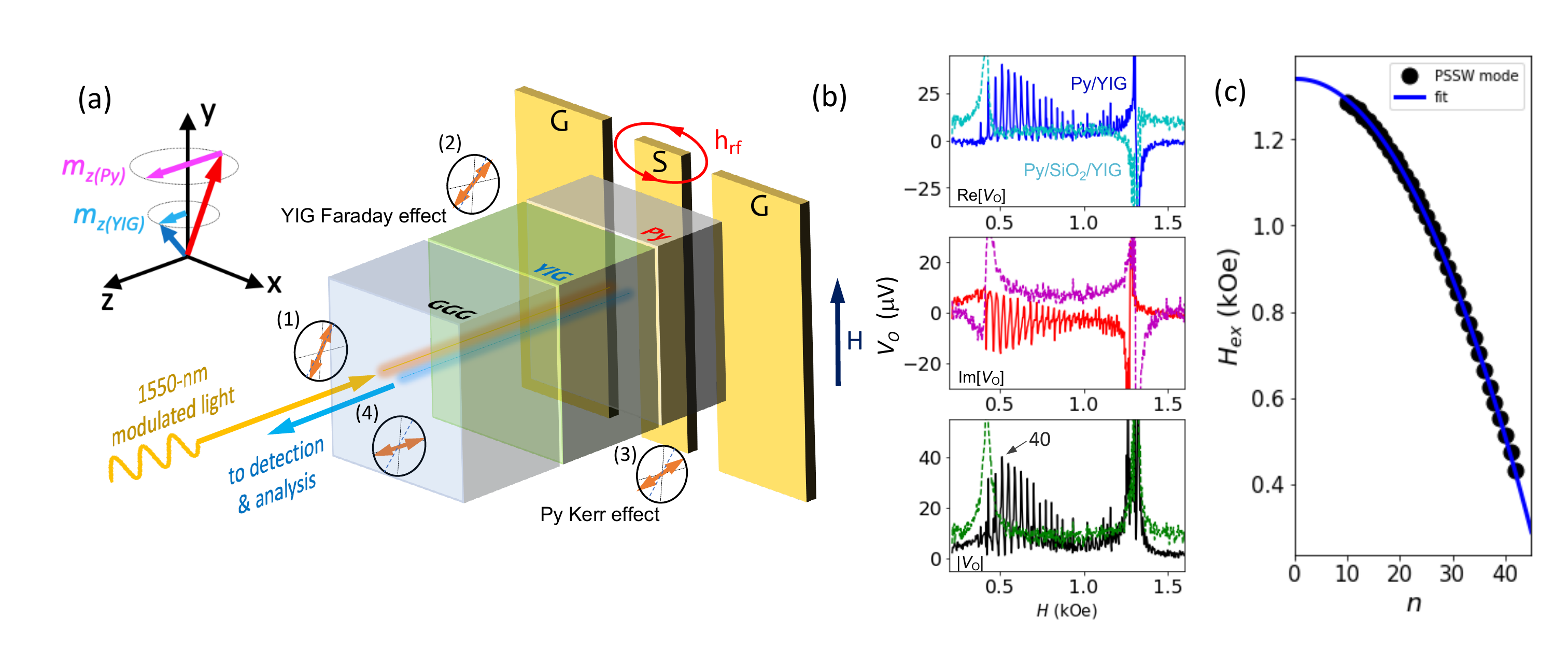}
 \caption{(a) Schematic illustration of the experimental setup. Modulated and linearly-polarized 1550-nm light enter the sample at a polarization angle (1); dynamic Faraday effect of the YIG causes the polarization to rotate (2); dynamic Kerr effect of the Py causes polarization to further rotate (3); the reflected light, upon the returning path, picks up again the Faraday effect and causes the polarization to further rotate (4), before entering light detection and analysis. The applied dc magnetic field is parallel to the ground-signal-ground (G-S-G) lines of the CPW. (b) \textcolor{black}{Example signal trace for YIG/Py (solid) and YIG/SiO$_2$/Py (dashed)} measured at 5.85 GHz, showing the in-phase $X$ (top) and quadrature $Y$ (middle), and the total amplitude, $\sqrt{X^2+Y^2}$ (bottom). (c) Plotting and the fitting of the observed PSSW modes versus the resonance fields.}
 \label{fig1}
\end{figure*}

In this work, we investigate the magnon-magnon coupling in YIG/Permalloy(Py) bilayers by a phase-resolved, heterodyne optical detection method. \textcolor{black}{We reveal the \textcolor{black}{coupled magnon modes} in the regime exhibiting the magnetically-induced transparency (MIT) effect, i.e. the magnetic analogy of electromagnetically induced transparency \textcolor{black}{(EIT) \cite{xufeng_prl2014,yuri_rmp2010,yuri_rmp2017,peng_ncomm2014,weis_science2010,naeini_nature2011}}, akin to a spin-wave induced suppression of FMR. \textcolor{black}{In the hybrid magnon-photon systems, the MIT effect arises when the coupling strength, $g/2\pi$ is larger than the photon dissipation rate $\kappa_\text{p}/2\pi$ but smaller than the magnon dissipation rate $\kappa_\text{m}/2\pi$ \cite{xufeng_prl2014}.} Under such a condition, the \textcolor{black}{mode hybridization} leads to an abrupt suppression of the microwave transmission at a certain frequency range. A transparency window, whose linewidth is determined by the low-loss mode, can be observed in the broad resonance of the other lossy mode. Such resonant transparency is controlled by an external magnetic field.} Our measurement is achieved via detecting the coupled magnetization dynamics of the insulating and metallic FMs using a single 1550-nm telecommunication wavelength. Unlike the ultrafast optical pump-probes \cite{rasing_rmp2010}, the method herein is a continue-wave (cw), \textcolor{black}{heterodyne} technique in which the 1550-nm laser light is modulated at the FMR frequencies (in GHz range) simultaneously with the sample's excitation. This feature makes the method effectively an optical ``lock-in" type measurement, akin to the electrical lock-in detection \cite{yoon_prb2016,yili_prap2019,yili_ieee2019}. The phase information between the Py and YIG FMRs, as well as the YIG perpendicular standing spin waves (PSSWs) can be obtained by simultaneously analyzing both the Kerr and Faraday responses.

\section{Samples and Measurements}

The commercial YIG films (from MTI Corporation) used in this work are 3-$\mu$m thick, single-sided grown on double-side-polished Gd$_3$Ga$_5$O$_{12}$ (GGG) substrates via liquid phase epitaxy (LPE). The Py films (t$_{Py}$ = 10 nm and 30 nm) were subsequently deposited on the YIG films using magnetron sputtering following earlier recipes \cite{yili_magnon2019}. To ensure the strong coupling, we used \textit{in situ} Ar gas rf-bias cleaning for 3 minutes, to clean the YIG surface before depositing the Py layer. Reference samples of \textcolor{black}{GGG/}YIG/SiO$_2$(3-nm)/Py(10-nm) \textcolor{black}{GGG/}YIG/Cu(3-nm)/Py(10-nm) were also prepared at the same growth condition.

Figure \ref{fig1}(a) illustrates the measurement configuration. The modulated and linearly-polarized 1550-nm light passes through the transparent GGG substrates and detects the dynamic Faraday and Kerr signals upon their FMR excitation. As the light travels through the YIG bulk, the dynamic Faraday rotation due to the YIG FMR is picked up. Similarly, the dynamic Kerr rotation caused by the Py FMR is then picked up, when the light reaches the Py layer. The Py layer also serves as a mirror and reflects the laser light. Upon reflection, the dynamic Faraday effect from the YIG is picked up again, making the effective YIG thickness 6-$\mu$m, i.e. twice the film thickness. It should be noted that the Faraday rotations for the incoming and returning light add up as opposed to cancel, due to the inversion of both the chirality of the Faraday rotation and the projection of the perpendicular magnetization of YIG along the wavenumber direction, whose mechanism is akin to a commercial ``Faraday rotator" often encountered in fiber optics.

The YIG/Py samples are chip-flipped on a coplanar waveguide (CPW) for microwave excitation and optical detection, as depicted in Fig. \ref{fig1}(a). An in-plane magnetic field, $H$, along the $y$-direction saturates both the YIG and Py magnetizations. We scan the frequency (from 4 to 8 GHz) and the magnetic field, and then measure the optical responses using a lock-in amplifier's \textcolor{black}{in-phase $X$ (Re[$V_\text{O}$]) and quadrature $Y$ (Im[$V_\text{O}$])} channels as well as the microwave transmission using a microwave diode. A detailed description of the measurement setup is in the Supplemental Materials (SM), Figure S1.

\section{Results and Discussions}

Figure \ref{fig1}(b) \textcolor{black}{compares the optical rectification signals between the 10-nm-Py sample and the YIG/SiO$_2$/Py reference sample} measured at 5.85 GHz. The 10-nm-Py sample (solid line) shows the representative features of the detected FMR and hybridized PSSW modes. The complete fine-scan including the FMR diode dataset are summarized in the SM, Fig. S2. The optical signals with the phase information are obtained by the lock-in's in-phase $X$ (Re[$V_\text{O}$], top panel) and quadrature $Y$ (Im[$V_\text{O}$], middle panel), \textcolor{black}{which are further used to calculate the total amplitude, $\sqrt{X^2+Y^2}$, (bottom panel). The technical details of the measured signal versus the optical and electrical phases are summarized in the SM.} 

\textcolor{black}{The YIG FMR signal at $\sim$1.3 kOe is accumulated from the Faraday effect corresponding to the spatially uniform precession of the YIG magnetization.} The FMR dispersion is described by the Kittel formula: $\omega^2/\gamma^2= H_\text{FMR}(H_\text{FMR}+M_s)$, where $\omega$ is the mode frequency, $\gamma/2\pi=(g_\text{eff}/2) \times$ 28 GHz/T is the gyromagnetic ratio, \textcolor{black}{$g_\text{eff}$ is the g-factor,} $H_\text{FMR}$ is the resonance field, and $M_s$ is the magnetization. \textcolor{black}{The excitation of the YIG PSSW modes introduces an additional exchange field $H_{ex}$ to the Kittel equation, as} $\mu_0 H_\text{ex} = (2A_\text{ex}/M_s) (n\pi/d_\mathrm{YIG})^2$, which defines the mode splitting between the PSSW modes and the uniform mode. Here $A_\text{ex}$ is the exchange stiffness, and $d_\mathrm{YIG}$ is the YIG film thickness. A total of more than 30 PSSW modes can be identified for the 10-nm-Py sample. \textcolor{black}{In Fig. \ref{fig1}(c), the quadratic increase of $H_\text{ex}$ with the mode number $n$ confirms the observation of the PSSWs. Fittings to the Kittel equation and the exchange field expression yield} $M_s$ = 1.97 kOe and $A_\text{ex}$ = \textcolor{black}{3.76} pJ/m, which are in good agreement with the previously reported values \cite{weiler_prl2018,yu_prl2018,qin_srep2018,yili_magnon2019}. 

\textcolor{black}{In Fig. \ref{fig1}(b), the Py FMR at $\sim$0.6 kOe is strongly modulated by the YIG PSSWs, exhibiting the MIT effect, due to the formation of hybrid magnon modes. Besides, the YIG PSSW signals near the Py FMR regime ($n>25$) are much stronger than the off-resonance regime ($n<25$), which indicates the important role of the Py/YIG coupling in exciting the relevant PSSW modes and resonantly enhancing the magnetization dynamics. As a comparison, no apparent PSSW modes are observed for the Py/SiO$_2$/YIG reference sample, in Fig.\ref{fig1}(b) (dashed line), indicating that only the Py but not the YIG PSSWs couples to the microwave drive in the MIT regime. The Py resonance linewidth also is much narrower.}

\begin{figure}[htb]
 \centering
 \includegraphics[width=4.3 in]{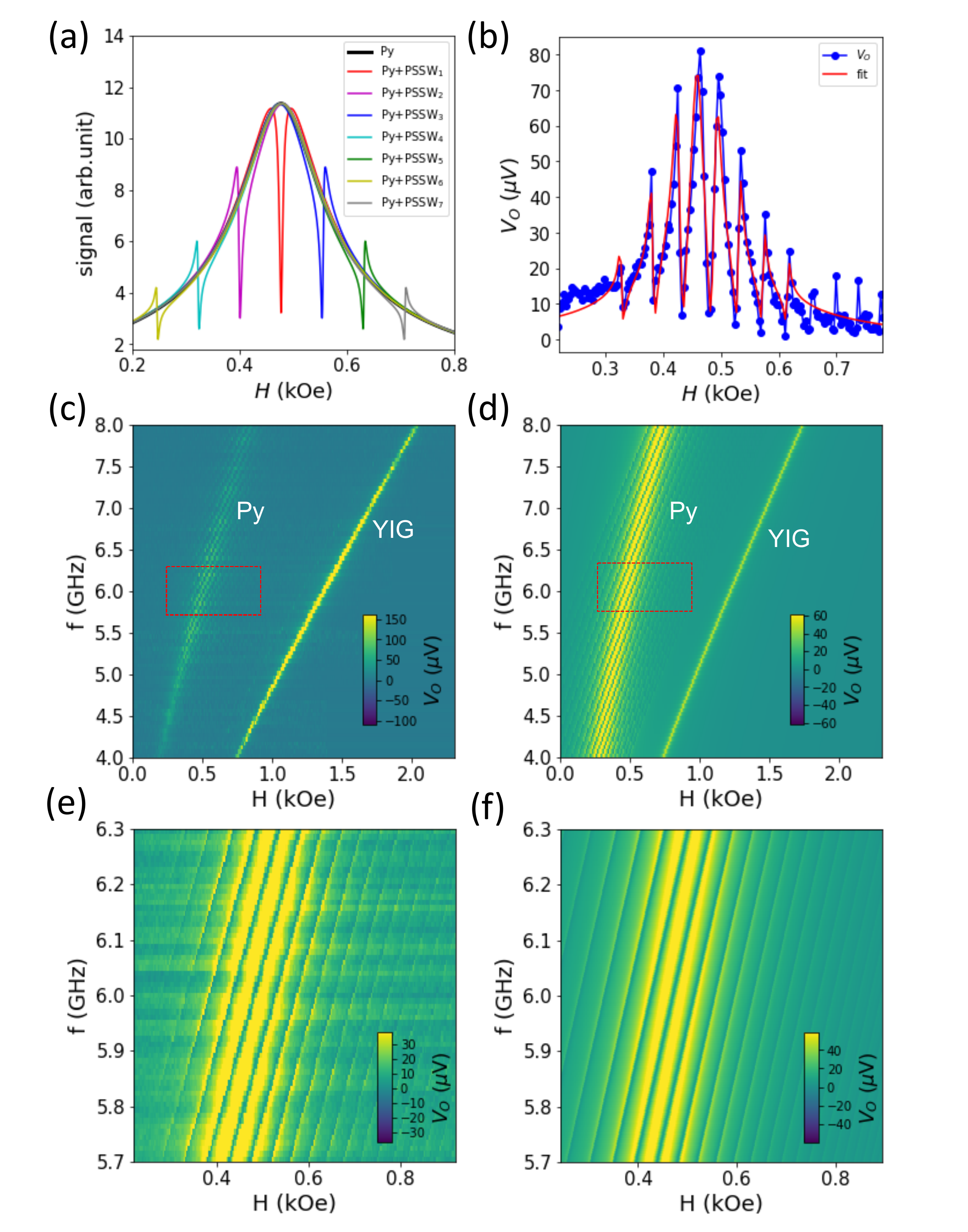}
 \caption{(a) \textcolor{black}{Theoretical signal trace of the MIT effect of the YIG/Py bilayer. 7 hybrid PSSW modes are shown as an example.} (b) \textcolor{black}{Example fitting of the complex optical singal at 6 GHz for the 30-nm-Py sample.} (c) \textcolor{black}{Full scan of the signal as a function of the magnetic field and frequency.} (d) \textcolor{black}{Theoretical calculated dispersion using the fitting parameters, reproducing the experimental data in (c).}  (e) and (f) \textcolor{black}{are the fine-scans at smaller field and frequency steps corresponding to the boxes in (c) and (d) (5.7 - 6.3 GHz).}}
 \label{fig2}
\end{figure}

\textcolor{black}{Our experimental configuration, similar to previously reported \cite{weiler_prl2018,yu_prl2018,qin_srep2018,yili_magnon2019}, is relevant to the Schl\"omann excitation mechanism of spin wave \cite{schlomann_jap1964} with a dynamic pinning at the interface \cite{wigen_prl1962}. The interfaces of two distinct, coupled magnetic layers have been recently recognized as an interesting source of spin dynamics generation and manipulation \cite{kruglyak_jphysd2017,verba_prb2020}. In particular, the critical role of the microwave susceptibilities of the distinct magnetic layers has been theoretically laid out \cite{poimanov_jphysd2019,poimanov_prb2018} that are directly relevant to the magnon-magnon coupled experiments \cite{weiler_prl2018,yu_prl2018,qin_srep2018,yili_magnon2019}.} Here, we introduce a phenomenological model by considering a series of YIG harmonic oscillators coupled with the Py oscillator, and using practical experimental fitting parameters, in which the measured complex optical signal, $V_\text{O}$, can be expressed as:
\begin{equation}
V_\text{O} = \frac{A e^{i(\phi_\text{L}-\phi_\text{m})}}{i(H^\text{Py}_\text{FMR} - H) - \Delta H_\text{Py} + \Sigma_j\frac{g^2}{i(H^\text{YIG}_{\text{PSSW},j} - H) - \Delta H_{\text{YIG},j}}}
\label{eq01}
\end{equation}
where $A$ is the total signal amplitude, $H_\text{FMR}^\mathrm{Py}$ and $H_\text{PSSW}^\mathrm{YIG}$ is the resonance field of Py and YIG-PSSWs, respectively, $\Delta H_\mathrm{YIG(Py)}$ is the half-width-half-maximum linewidth, and $g$ is the \textcolor{black}{fieldlike coupling strength from the interfacial exchange}. \textcolor{black}{ This model disregards the linear frequency-dependent phase (exists in the Re[$V_\text{O}$] and Im[$V_\text{O}$] signals) but directly analyzes the total optical signal.} 

\textcolor{black}{Figure \ref{fig2}(a) plots the theoretically predicated MIT effect according to Eq.\ref{eq01}, showing the lineshape of the Py FMR mode that is coupled to the YIG PSSW modes. The center curve with a zero resonance detuning denotes the MIT effect. \textcolor{black}{The magnon-magnon coupling induces a set of sharp dips in the spectra. Such dips in the optical reflection means a peak in their transmission, which is referred to as a ``transparency window" in quantum optics, resembling the EIT phenomenon in photonics \cite{peng_ncomm2014} and optomechanics \cite{weis_science2010,naeini_nature2011}.} Using a single $g$ value, the amplitude and phase of the hybrid modes display a clearly evolution with respect to the different Py-YIG resonance detuning. Away from the Py FMR, the lineshape appears to be more antisymmetric, whilst around the Py FMR, the hybrid mode appears to be more symmetric. Such a phase evolution is contained in the Eq.\ref{eq01} and is not a fit parameter. Fig.\ref{fig2}(b) is an example fitting result of a signal trace at 6 GHz.} The fittings nicely reproduce the complex lineshapes arising from the coupled YIG PSSW modes and the Py FMR, including the phase evolution across the involved PSSW modes ($n$). \textcolor{black}{Our model allows extracting the YIG and Py magnon dissipation rates, $\Delta H_\text{YIG}$ = 1.8 Oe, $\Delta H_\text{Py}$ = 43.3 Oe, and the coupling strength: $g$ = 18.7 Oe. In the frequency domain, these values correspond to $g/2\pi = 90$ MHz, $\kappa_\text{YIG}/2\pi = 6$ MHz, and $\kappa_\text{Py}/2\pi = 308$ MHz. The numbers satisfy the condition for the MIT effect: $\Delta H_\text{YIG} < g < \Delta H_\text{Py}$.}  

\textcolor{black}{The YIG/Py interfacial exchange coupling can be also found from the $H^\text{Py}_\text{FMR}$ shift comparing to the YIG/SiO$_2$/Py reference sample. As shown in Fig. \ref{fig1}(b), the Py resonance occurs at a higher field when Py is in direct contact with YIG due to the interfacial exchange coupling. The increase of $H^\text{Py}_\text{FMR}$ suggests that the YIG/Py interface induces a negative effective field onto Py, which agrees well with the antiferromagnetic coupling in the previous reports \cite{weiler_prl2018,yili_magnon2019}. We find a resonance offset of $H^\text{Py}_\text{FMR,ofst}=0.17$ kOe from Fig. \ref{fig1}(b), which further yields a fieldlike coupling strength from the interfacial exchange \cite{yili_magnon2019}, $g =H^\text{Py}_\text{FMR,ofst}\times 0.9\sqrt{M_\text{Py}t_\text{Py}/M_\text{YIG}t_\text{YIG}}$. Taking $M_\text{YIG}=1.97$ kOe, $M_\text{Py}=10$ kOe, $t_\text{Py}=10$ nm and $t_\text{YIG}=3$ $\mu$m, we estimate $g = 19.8$ Oe. This value is in good agreement with that obtained earlier from the MIT effect.}

\textcolor{black}{Figure \ref{fig2} (c and d) compare the experimental and theoretical spin-wave dispersion using the total amplitude signal ($\sqrt{X^2+Y^2}$), whilst the individual channels, $X$ and $Y$, as well as the corresponding theoretical plots, are included in the SM, Fig. S3.} To better analyze the hybrid PSSW modes, we show the zoom-in scan between 5.7 and 6.3 GHz and 0.2 to 0.9 kOe in Fig.\ref{fig2} (e and f), which covers the Kittel dispersion of the Py. We clearly identify the distinct PSSW modes strongly ``chopping" the Py FMR line. In particular, the Py resonance is attenuated to nearly the background level (non-absorption condition) at the PSSW resonance dips.    

The same measurements are also performed for the reference samples, YIG/SiO$_2$/Py and YIG/Cu/Py, as summarized in the SM, Fig. S4. Despite the observation of the YIG and Py FMR modes, we do not measure any signal of the \textcolor{black}{hybrid PSSW modes}, \textcolor{black}{reconfirming that the excitation of the PSSW modes are primarily via the interfacial exchange coupling, rather than the dipolar interactions.}  

 \begin{figure}[htb]
 \centering
 \includegraphics[width=5 in]{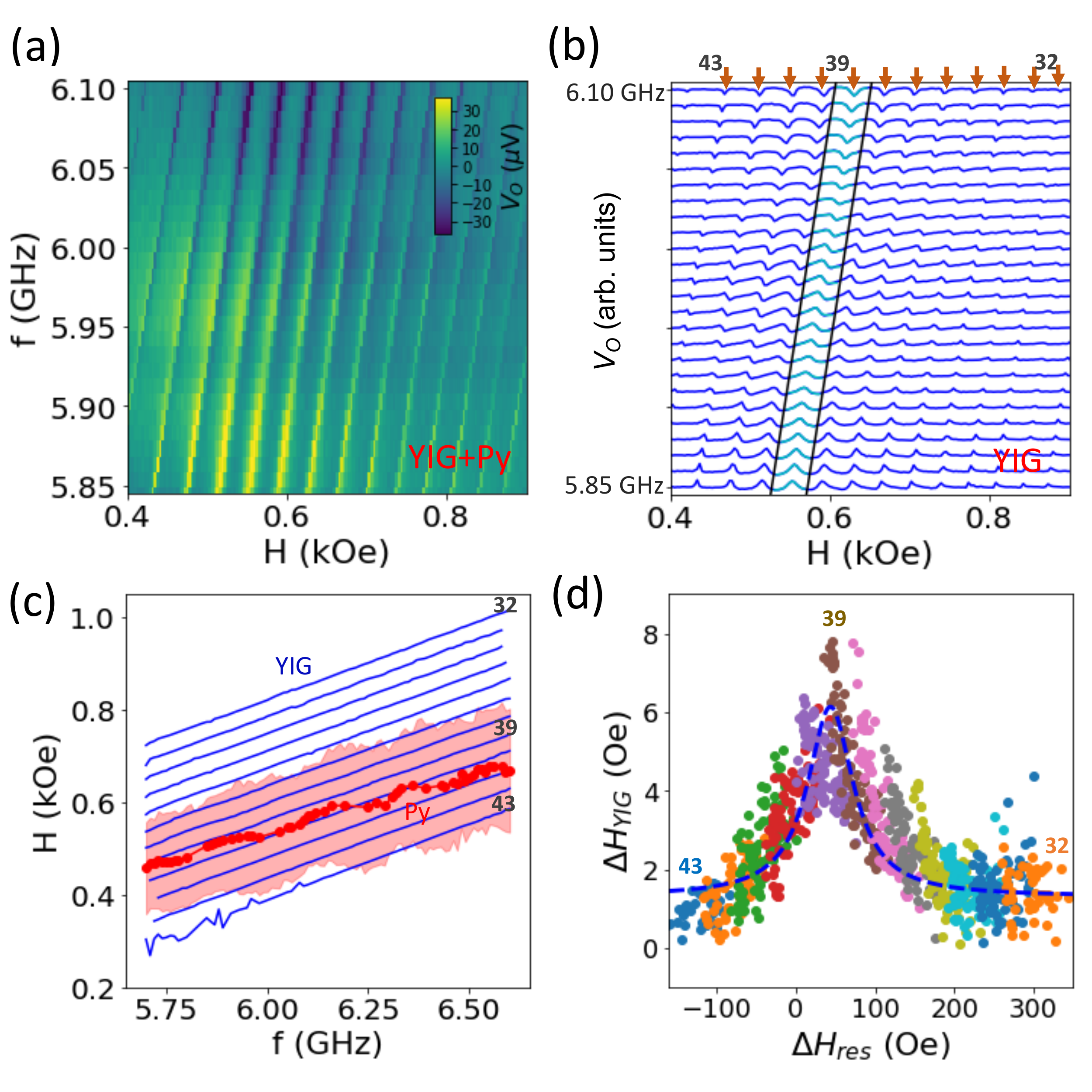}
 \caption{(a) \textcolor{black}{The Re[$V_\text{O}$] signal at a representative frequency window (5.85 - 6.1 GHz) for the 10-nm-Py sample.} (b) \textcolor{black}{YIG PSSW lineshape (12 mode series near the Py resonance are labeled and analyzed, $n$ = 32 - 43) after subtracting the Py resonance profile. The highlighted section is an example series at $n$ = 39.} (c) \textcolor{black}{Resonance field, $H^\text{YIG}_\text{PSSW}$ of the PSSW series and the $H^\text{Py}_\text{FMR}$ envelope. The shaded area reflects the Py linewidth.} (d) \textcolor{black}{The extracted YIG PSSW linewidth $\Delta H_\text{YIG}$ versus the $\Delta H_\text{res}$ at each frequency and for all the PSSW series.}\textcolor{blue}{} }
 \label{fig3}
\end{figure}

\textcolor{black}{To further examine the detuning range and its characteristics, we separate the Py resonance envelope with the YIG PSSW modes. Such analysis can be made via fitting either the raw Re[$V_\text{O}$] or Im[$V_\text{O}$] data.} \textcolor{black}{Fig. \ref{fig3}(a) shows the raw Re[$V_\text{O}$] signal of the hybrid modes at a representative frequency window (5.85 - 6.1 GHz) for the 10-nm-Py sample. After subtracting the Py resonance profile (details are in the SM, Fig. S5), we can fit each PSSW series (labeled $n = 32 - 43$) to a phase-shifted Lorentzian function yielding the resonance and linewidth for each PSSWs, as shown in Fig.\ref{fig3}(b) (where the highlighted section shows an example series at $n = 39$).} We define a ``resonance distance", $\Delta H_\text{res}$ = $ [H^\text{YIG}_\text{PSSW} - H^\text{Py}_\text{FMR}]$, which represents their frequency detuning and the coupling efficiency.

\textcolor{black}{Figure \ref{fig3}(c) shows the resonance field $H^\text{YIG}_\text{PSSW}$ of the PSSW series (thin lines, from $n = 32 - 43$) comparing to the $H^\text{Py}_\text{FMR}$ (single thick line). The shaded area indicates the Py linewidth, which is centered at the $H^\text{Py}_\text{FMR}$ and is also much enhanced as compared to the case without the mode coupling (in the YIG/SiO$_2$/Py sample). Next, we plot the $\Delta H_\text{res}$ at each frequency and for all the PSSW series with the corresponding YIG PSSW linewidth, $\Delta H_\text{YIG}$, in Fig. \ref{fig3}(d).} We clearly observe a \textcolor{black}{modulation effect} of the YIG linewidth, $\Delta H_\text{YIG}$, from $\sim$ 2 Oe to $\sim$ 10 Oe, spanning across the magnon-magnon coupling regime. This observation provides strong evidence that the MIT linewidth is broadened due to the additional energy dissipation by coupling the YIG PSSW modes to the Py FMR mode, also known as the Purcell regime \cite{hu_ssp2018,bhoi_ssp2019,xufeng_prl2014}. \textcolor{black}{From the theoretical model in Eq.\ref{eq01}, we  obtain a relationship between the YIG linewidth broadening due to a finite $g$ and the overlapped resonance:
\begin{equation}
\Delta H^\text{MIT}_{\text{YIG},j}  = \Delta H_{\text{YIG},j} + g^2 \frac{\Delta H_\text{Py}}{(\Delta H_\text{res})^2 + (\Delta H_\text{Py})^2}
\label{eqs02}
\end{equation}
where the derivation is included in the SM. Since the MIT regime is within a relatively narrow frequency window from 5.7 - 6.6 GHz, we take the average of the Py linewidth within this frequency window, as $\Delta H_\text{Py} = 60 \pm 8 $ Oe from Fig.\ref{fig3}(c). For YIG, we take the same linewidth as in Fig.\ref{fig2}(b), $\Delta H_{\text{YIG},j} = 1.8$ Oe, therefore leaving $g$ as the only fitting parameter. The best fit yields $g = 17.7 \pm 1.2$ Oe, with a fitting curve indicated in Fig.\ref{fig3}(d), dashed line}. This value agrees with the lineshape fitting results of 18.7 Oe as discussed above.  

\textcolor{black}{Finally, the observed multiple PSSW modes and their coupling to a ``magnonic cavity" are similar to the multi-mode coupling in the magnon-photon system \cite{xufeng_jap2016}, in which the profiles and properties of each PSSW mode are greatly modified as compared to the free-space conditions. We envisage that a stronger coupling condition may be fulfilled by combining an appropriately designed optical cavity such as the whispering gallery modes \cite{xzhang_prl2016}, or replacing the Py with a low damping ferromagnetic material with reduced dissipation rate \cite{weiler_prl2018}. }

\section{Conclusion}

In summary, we report the observation of the magnetically-induced transparency in YIG/Py bilayers exhibiting magnon-magnon coupling. The use of the thin-film YIG system shows great potential in practical applications. The series of standing waves in YIG may allow to build an evenly distributed resonance array in a single YIG device, which may lead to relevant applications such as memory and comb generation. In addition, compared with the so-far widely used hybrid magnonic systems that utilize the ferromagnetic resonances, our results pave the way towards building more complex hybrid systems with spin-waves. Our  measurement  is  achieved  via  a simultaneous and stroboscopic detection of the coupled  magnetization dynamics using a single wavelength, therefore avoids the possible artifacts due to multiple probes. Our work, performed in a planar structure as opposed to 3D cavities, also paves the way towards solving strong magnon-magnon couplings by the state-of-the-art spin-orbitronic toolkits \cite{wz_jap2015,hyunsoo_apr2018}, involving emerging materials such as antiferromagnets \cite{baltz_rmp2018,wz_mser2016,lukas_prl2019}, 2D monolayers \cite{macneill_nphys2017,wz_aplm2016,yili_ieeetted2018,xzhang_amse}, and topological insulators \cite{mellnik_nature2014,peng_sciadv2019}.  \newline

\textbf{Acknowledgements}\\ 
W.Z. acknowledges useful discussions with V. Tyberkevych and A. Slavin. This work, including apparatus buildup, experimental measurements, and data analysis, was supported by AFOSR under Grant No. FA9550-19-1-0254, National Science Foundation under Grants No. DMR-1808892 and ECCS-1933301 . Work at Argonne, including sample preparation, was supported by  U.S. DOE, Office of Science, Materials Sciences and Engineering Division. M.H. acknowledges the Michigan Space Grant Consortium Student Fellowship for financial support.\\  

\textbf{Author Contributions}\\ 
Y.L. and W.Z. conceived the idea. Y.X., M.H., R.B., W.Z. performed the experiment. J.S., X.Z., and Y.L. developed the theoretical model. J.P., A.H., and V.N. participated in the thin-film sample fabrication. G.S. and H.Q. participated in the microwave measurement. Y.L., X.Z., J.S., and W.Z. prepared the figures and manuscript draft. All authors discussed the results and participated in the writing and finalizing of the manuscript.\\   

\textbf{Competing Interests}\\ 
Te authors declare no competing interests.\\

\textbf{Additional Information}\\ 
\textbf{Supplementary information} is available for this paper at https://doi.org/XX.XXXX.\\
\textbf{Correspondence} and requests for materials should be addressed to Y.L. and W.Z.\\

\end{document}